\documentstyle[aps,prl,epsfig,floats]{revtex}

\tighten

\newcommand{\tr}{\mbox{Tr}\,}
                 
\newcommand{\st}{{\scriptscriptstyle T}}

\newcommand{\bm}{\bbox}

\newcommand{\dtslash}{{\bm D}\hspace{-0.23cm}/_{\st}}
\newcommand{\npslash}{n\hspace{-0.18cm}/_{+}}
\newcommand{\nmslash}{n\hspace{-0.18cm}/_{-}}
\newcommand{\stslash}{S\hspace{-0.20cm}/_{\st}}
\newcommand{\atslash}{{\bm A}\hspace{-0.21cm}/_{\st}}
\newcommand{\deltslash}{{\bm \partial}\hspace{-0.20cm}/}

\newcommand{\dbar}{\; \overline{\phantom{I}} \hspace{-0.25cm} D}
\newcommand{\kbar}{\; \overline{\phantom{I}} \hspace{-0.25cm} K}
\newcommand{\obar}{\; \overline{\phantom{I}} \hspace{-0.24cm} O}

\begin{document}
%%%%%%%%%%%%%%%%%%%%%%%%%%%%%%%%%%%%%%%%%%%%%%%%%%%%%%%%%
 
%@twocolumnfalse\endcsname

\title{
\begin{flushright}
\begin{minipage}{4 cm}
\small
%VUTH 01-??\\
\end{minipage}
\end{flushright}
Higher twist and transverse momentum dependent parton distributions: 
a light-front hamiltonian approach} 

\author{R.~Kundu and A.~Metz}

\address{
Division of Physics and Astronomy, Faculty of Science, Free University \\
De Boelelaan 1081, NL-1081 HV Amsterdam, the Netherlands\\[2mm]}

%\date{draft of \today}
\date{\today}

\maketitle

%%%%%%%%%%%%%%%%%%%%%%%%%%%%%%%%%%%%%%%%%%%%%%%%%%%%%%%%%%%%%%%%%%%%%%%%%%%
\begin{abstract}

In order to study twist-3 and transverse momentum dependent parton distributions, 
we use light-front time-ordered pQCD at order $\alpha_s$ to calculate various 
distribution functions for a dressed quark target.
This study enables us to investigate in detail the existing relations between 
twist-3 and transverse momentum dependent parton distributions.
Our calculation shows explicitly that two versions of such relations,
considered to be equivalent, occur in the 
literature which need to be distinguished.
Moreover, we examine sum rules for higher twist distributions.
While the Burkhardt-Cottingham sum rule for $g_2$ is fulfilled, the corresponding 
sum rule for $h_2$ is violated.
\end{abstract}

\pacs{13.60.Le,13.87.Fh,12.39.Fe}

%%%%%%%%%%%%%%%%%%%%%%%%%%%%%%%%%%%%%%%%%%%%%%%%%%%%%%%%%%%%%%%%%%%%%%%%%%%
\section{Introduction}
%%%%%%%%%%%%%%%%%%%%%%%%%%%%%%%%%%%%%%%%%%%%%%%%%%%%%%%%%%%%%%%%%%%%%%%%%%%

In view of the increasing accuracy of recent and planned high energy scattering
experiments, more and more attention is paid to the study of parton distributions which 
are of higher twist and (or) dependent on the transverse momenta of the partons. 
The twist-3 distribution functions are accessible through the measurement of certain 
asymmetries in polarized DIS \cite{abe_96} and Drell-Yan processes \cite{jaffe_91}.
The transverse momentum dependent ($k_{\st}$-dependent) structure functions play an
important role both in Drell-Yan processes and semi-inclusive DIS \cite{ralston_79}.
In such reactions, e.g., the transverse momenta and the transverse spin of the partons can get 
coupled giving rise to azimuthal asymmetries 
(see e.g. Refs. \cite{collins_93,mulders_96,boer_98,levelt_94}), 
which are very suitable observables for studying the correlations of quarks and gluons in 
hadrons.
Often, effects due to higher twist and transverse momenta appear simultaneously like
in the recent HERMES measurements of the longitudinal single spin asymmetry in semi-inclusive
pion production \cite{hermes_00}.
In this work, we study these higher twist and $k_{\st}$-dependent structure functions and
their interrelations in the framework of light-front hamiltonian QCD.

As is well-known, twist-3 and $k_{\st}$-dependent parton distributions are 
related \cite{mulders_96,boer_98,bukhvostov_84,belitsky_97a,belitsky_97b} as a consequence of 
Lorentz invariance.
These relations impose important constraints on the distribution functions, which allow
one to eliminate unknown structure functions in favor of known ones whenever applicable.
Consequently, they have been used frequently in the literature to facilitate matters, for 
instance in studying the evolution of $k_{\st}$-dependent distribution 
functions \cite{bukhvostov_84,belitsky_97a,henneman_01}. 

Our motivation here is to investigate the validity of these Lorentz invariance relations by 
explict calculation of all the involved distribution functions.
There exists a very convenient tool based on the light-front hamiltonian description of composite 
systems utilizing many-body wave-functions, which enables us to study these relations in the 
context of perturbative QCD. 
This tool has already been used successfully in the literature to calculate unpolarized and 
polarized parton distributions \cite{harindranath_98} as well as the transversity 
distribution \cite{mukherjee_01}.
The simplicity of this approach has also been exploited to make a critical examination of 
the Wandzura-Wilczek relation \cite{harindranath_97}. 
Here, we use the same approach to calculate the higher twist and $k_{\st}$-dependent parton 
distributions perturbatively to order $\alpha_s$, and then study the Lorentz invariance 
relations by employing a dressed quark target.
We demonstrate that there exist two sets of relations, although assumed to be the same, 
are not identical in reality.
More precisely, only one set of relations is verified for a dressed quark target, 
whereas the drawback in the other case can be traced back to the absence of quark-gluon-quark
correlators, which seem to be crucial ingredients in the Lorentz invariance relations in 
a gauge theory.

Our calculation also gives us the opportunity to investigate the sum rules for the
twist-3 distributions $g_2$ and $h_2$. 
The Burkhardt-Cottingham (BC) sum rule for $g_2$ \cite{burkhardt_70} is satisfied for the 
dressed quark target, but the corresponding sum rule for $h_2$ \cite{burkardt_92,tangerman_94} 
turns out to be violated.
To our best knowledge, the violation of the sum rule for $h_2$ in a perturbative treatment is
a new observation.
 
The paper is organized as follows: In Sec.~\ref{sec_2}, we outline the definition of the parton
distributions relevant for our discussion, and give a detailed account of their relations due
to Lorentz invariance.
In Sec.~\ref{sec_3}, first the operators for twist-3 distributions are expressed in terms of 
dynamical fields, and some relevant points regarding the dressed quark target are discussed.
Then we present our results for the different parton distributions and a detailed 
investigation of the two sets of Lorentz invariance relations. 
In Sec.~\ref{sec_4}, we study the sum rules for $g_2$ and $h_2$ and conclude in 
Sec.\ref{sec_5}. 
Some conventions are summarized in an Appendix.

%%%%%%%%%%%%%%%%%%%%%%%%%%%%%%%%%%%%%%%%%%%%%%%%%%%%%%%%%%%%%%%%%%%%%%%%%%%
\section{Parton Distributions: Definitions and their Interrelations} \label{sec_2}
%%%%%%%%%%%%%%%%%%%%%%%%%%%%%%%%%%%%%%%%%%%%%%%%%%%%%%%%%%%%%%%%%%%%%%%%%%%
In this section we recall the definitions of various parton distributions that already 
exist in the literature and introduce the Lorentz invariance relations among them. 
We restrict the discussion below to twist-3 structure functions, while in the case of 
$k_\st$-dependent functions we limit ourselves to twist-2 level which is sufficient for 
our purpose. 
For a complete discussion one should go back to the original references mentioned below.  

To begin with, we specify the correlator $\Phi(x)$ of two quark fields on the 
light-front\footnote{Our definition of light-front components of a generic 4-vector
as well as further conventions are summarized in the Appendix.} in terms of
which all the structure functions are defined,
\begin{equation} \label{correl}
\Phi_{ij}(x) = \frac{1}{\sqrt{2}} \int \frac{d\xi^-}{2\pi} \, e^{i k \cdot \xi} \,
\langle P,S | \; \bar{\psi}_{j}(0) \, \psi_{i}(\xi) \; | P,S \rangle 
\bigg|_{\xi^+ = {\bm \xi}_{\st} = 0} \,,
\end{equation}
with  $k^+ = x P^+$. 
The target state is characterized by its four-momentum $P$ and the covariant spin vector
$S$ $(P^2 = M^2, \, S^2 = -1, \, P \cdot S = 0)$. 
Note that $\Phi(x)$ can easily be made gauge invariant by putting an appropriate gauge 
link between the quark fields. 
However, since the non-localilty in the operator is only in the longitudinal direction $\xi^-$ 
and we shall be working in the light-front gauge $(A^+ = 0)$, we can always get rid off the 
gauge link in Eq.~(\ref{correl}).

Now, the parton distributions appear in a general decomposition of the correlator $\Phi(x)$ 
where one finds three functions at twist-2 and three functions at twist-3 
level \cite{jaffe_91,mulders_96},
\begin{eqnarray} \label{phi_decomp}
\Phi(x) & = & \frac{1}{2} \, \bigg\{ f_1(x) \, \npslash 
 + \lambda \, g_1(x) \, \gamma_5 \, \npslash
 + h_1(x) \, \gamma_5 \, \stslash \, \npslash \bigg\}
 \nonumber \\
 & & + \frac{M}{\sqrt{2} \, P^+} \, \bigg\{ e(x) + g_T(x) \, \gamma_5 \, 
 \stslash 
    + \lambda \, h_L(x) \, \gamma_5 \, \frac{[\npslash,\nmslash]}{2} \bigg\} \,,
\end{eqnarray}
with $n_+$ and $n_-$ being two light like vectors satisfying $n_+\cdot n_-=1$. 
The helicity of the target state is given by $\lambda$, while
$S_{\st}^{\mu} \equiv (0,0,{\bm S}_{\st})$ represents the transverse spin of the target.
Sometimes different notations for twist-2 distributions are used in the literature
$(f_1(x) = q(x), \; g_1(x) = \Delta q(x),\; h_1(x) = \Delta_{\st}q(x) = \delta q(x))$.
The twist-3 part contains the well-known transversely polarized structure function $g_T$, and
two chiral-odd distributions $e$ and $h_L$. 
Note that in Eq.~(\ref{phi_decomp}) we have not considered the so called T-odd parton 
distributions. 

The structure functions in Eq.~(\ref{phi_decomp}) are projected out by performing 
traces of $\Phi(x)$ with suitable Dirac matrices.
Using the abbreviation $\Phi^{[\Gamma]} \equiv \tr (\Phi \Gamma) / 2$, we give the 
explicit expressions for those structure functions that are relevant for our discussion 
here,
\begin{eqnarray}
 g_1(x) & = & \frac{1}{\sqrt{2} \, \lambda} \, \Phi^{[\gamma^+ \gamma_5]}
   = \frac{1}{4 \, \lambda} \int {d\xi^- \over 2\pi} \, e^{i k \cdot \xi} \,
     \langle P,S | \; \bar{\psi}(0) \, \gamma^+ \gamma_5 \, \psi(\xi) \; |P,S \rangle \,, 
     \label{g1phi} \\
 h_1(x) & = & \frac{1}{\sqrt{2} \, S_{\st}^i} \, \Phi^{[i \sigma^{i+} \gamma_5]}
   = \frac{1}{4 \, S_{\st}^i} \int {d\xi^- \over 2\pi} \, e^{i k \cdot \xi} \, 
     \langle P,S | \; \bar{\psi}(0) \, i\sigma^{i+} \gamma_5 \, \psi(\xi) \; |P,S \rangle \,, 
     \label{h1phi} \\
 g_T(x) & = & \frac{P^+}{\sqrt{2} \, M \, S_{\st}^i} \, \Phi^{[\gamma^i \gamma_5]}
   = \frac{P^+}{4 \, M \, S_{\st}^i } \int {d\xi^- \over 2\pi} \, e^{i k \cdot \xi} \, 
     \langle P,S | \; \bar{\psi}(0) \, \gamma^i \gamma_5 \, \psi(\xi) \; |P,S \rangle \,, 
     \label{gtphi} \\
 h_L(x) & = & \frac{P^+}{2 \sqrt{2} \, M \, \lambda} \, \Phi^{[i \sigma^{+-} \gamma_5]}
   = \frac{P^+}{8 \, M \, \lambda } \int {d\xi^- \over 2\pi} \, e^{i k \cdot \xi} \, 
     \langle P,S | \; \bar{\psi}(0) \, i\sigma^{+-} \gamma_5 \, \psi(\xi) \; |P,S \rangle \,,
     \label{hlphi} 
\end{eqnarray}
where, like in Eq.~(\ref{correl}), all the correlators are understood to be on the
light-front, i.e. $\xi^+ = {\bm \xi}_{\st} = 0$.

In a similar way, $k_\st$-dependent parton distributions are defined starting from the 
following correlation function where the non-locality in its operator structure is not 
only in $\xi^-$ but in ${\bm \xi}_\st$ as well,
\begin{equation} \label{correlkt}
\Phi_{ij}(x,{\bm k}_{\st}) = \frac{1}{\sqrt{2}} \int \frac{d\xi^- d^2 {\bm \xi}_{\st}}{(2\pi)^3} 
\, e^{i k \cdot \xi} \, \langle P,S | \; \bar{\psi}_{j}(0) \, \psi_{i}(\xi) \;
| P,S \rangle \bigg|_{\xi^+ = 0} \,.
\end{equation}
Here we have assumed that in $A^+=0$ gauge together with antisymmetric boundary conditions
for the transverse gluon field, the gauge link can still be omitted as argued, e.g., in 
Ref. \cite{mulders_96}.  
In the general decomposition of this correlator one naturally finds more distribution 
functions due to the presence of an extra vector $k_\st^\mu$~\cite{mulders_96},
\begin{eqnarray}
\Phi(x,{\bm k}_{\st}) & = & \frac{1}{2} \bigg\{ f_1(x,{\bm k}_{\st}^2) \, \npslash 
 + \bigg( \lambda \, g_{1L}(x,{\bm k}_{\st}^2) 
          + \frac{{\bm k}_{\st} \cdot {\bm S}_{\st}}{M} \, g_{1T}(x,{\bm k}_{\st}^2) \bigg) 
          \, \gamma_5 \, \npslash 
 \nonumber \\
 & & \hspace{0.5cm} 
 - h_{1T}(x,{\bm k}_{\st}^2) \, i \sigma_{\mu\nu} \, \gamma_{5} \, S_{\st}^{\mu} \, n_{+}^{\nu}
 - \bigg( \lambda \, h_{1L}^{\perp}(x,{\bm k}_{\st}^2) 
          + \frac{{\bm k}_{\st} \cdot {\bm S}_{\st}}{M} \, h_{1T}^{\perp}(x,{\bm k}_{\st}^2)\bigg) 
          \frac{i \sigma_{\mu\nu} \, \gamma_5 \, k_{\st}^{\mu} \, n_{+}^{\nu}}{M} \bigg\} \,.
\label{ktdepend}
\end{eqnarray}
Here we have shown only the twist-2 part, which is sufficient for our purpose and have omitted 
the T-odd functions as before.

Like in the previous case, one projects out the structure functions in Eq.~(\ref{ktdepend}) 
by performing traces of $\Phi(x, {\bm k}_{\st})$ with suitable Dirac matrices. 
Two of the projections necessary for our
discussion are,
\begin{eqnarray}
\frac{1}{\sqrt{2}} \, \Phi^{[\gamma^+ \gamma_5]}(x,{\bm k}_{\st}) & = &
 \lambda \, g_{1L}(x,{\bm k}_{\st}^2)
 + \frac{{\bm k}_{\st} \cdot {\bm S}_{\st}}{M} \, g_{1T}(x,{\bm k}_{\st}^2) \,,
\label{g1t1}
\\
\frac{1}{\sqrt{2}} \, \Phi^{[i \sigma^{i+} \gamma_5]}(x,{\bm k}_{\st}) & = &
 S_{\st}^i \, h_{1T}(x,{\bm k}_{\st}^2)
 + \frac{k_{\st}^i}{M} \bigg( \lambda \, h_{1L}^{\perp}(x,{\bm k}_{\st}^2) 
   + \frac{{\bm k}_{\st} \cdot {\bm S}_{\st}}{M} \, h_{1T}^{\perp}(x,{\bm k}_{\st}^2)\bigg) \,.
\label{h1p1}
\end{eqnarray} 
Note that depending on the target polarization the same projection of $\Phi(x,{\bm k}_{\st})$ 
allows one to calculate different structure functions. 
For example, from Eq.~(\ref{g1t1}) we get $g_{1L}(x,{\bm k}_\st)$ or $g_{1T}(x,{\bm k}_\st)$ 
for the target being polarized in longitudinal or transverse direction, respectively. 
Therefore, Eqs.~(\ref{correlkt}-\ref{h1p1}) give us a well-defined way to calculate 
$g_{1T}(x,{\bm k}_\st)$ and $h_{1L}^{\perp}(x,{\bm k}_\st)$ which are necessary for the
subsequent discussion. 
In what follows we need the $k_{\st}^2$-moments of these two functions, which are  
defined as~\cite{mulders_96}
\begin{equation} \label{moments}
g_{1T}^{(1)}(x) = \int d^2{\bm k}_{\st} \, \frac{{\bm k}_{\st}^2}{2 M^2}
  \, g_{1T}(x,{\bm k}_{\st}^2)\,, \qquad
h_{1L}^{\perp (1)}(x) = \int d^2{\bm k}_{\st} \, \frac{{\bm k}_{\st}^2}{2 M^2}
  \, h_{1L}^{\perp}(x,{\bm k}_{\st}^2) \,.
\end{equation}

Now that we have given all the definitions of relevant structure functions, we are 
in a position to discuss the existing relations among them.
These are usually of two kinds -- one follows from the QCD equations of motion and the 
other comes as a consequence of Lorentz invariance. 
Here we are mainly interested in the latter ones which, according to 
Refs.~\cite{mulders_96,boer_98}, read as
\begin{eqnarray} \label{rel_1a}
g_T(x) & = & g_1(x) + \frac{d}{dx} \, g_{1T}^{(1)}(x) \,,
 \\ \label{rel_1b}
h_L(x) & = & h_1(x) - \frac{d}{dx} \, h_{1L}^{\perp (1)}(x) \,.
\end{eqnarray}
These relations have been derived from the general Lorentz covariant decomposition 
of the correlation function $\Phi$ of two quark fields before it is constrained on the 
light-cone and, hence, they are quite naturally referred to as Lorentz invariance relations. 
On the other hand, a similar relation for $g_T$ also attributed to Lorentz-invariance 
has already been proposed in Ref.~\cite{bukhvostov_84} and extended for $h_L$ in 
Ref.~\cite{belitsky_97a}. 
A detailed account on these relations can be found in Ref. \cite{belitsky_97b}. 
Using the notations of Ref.~\cite{belitsky_97b}, with necessary modifications for the 
conventions that we follow, the corresponding relations are given by
\begin{eqnarray} \label{rel_2a}
g_{T}(x) & = & g_1(x) + \frac{d}{dx} \kbar(x) 
 + \int dx' \frac{\dbar(x,x') + \dbar(x',x)}{x' - x} \,,
 \\
\label{rel_2b}
h_{L}(x) & = & h_1(x) + \frac{1}{2} \frac{d}{dx} \widetilde{K}(x) 
 + \frac{1}{2} \int dx' \frac{\widetilde{D}(x,x') + \widetilde{D}(x',x)}{x' - x} \,.
\end{eqnarray}
Note that here quark-gluon-quark light-front correlators are involved, which depend on 
two momentum fractions denoted as $x={k^+\over P^+}$, $x'={k'^+\over P^+}$ and the new 
correlation functions appearing in Eq. (\ref{rel_2a}) are given by
\begin{eqnarray}
\kbar(x) & = & \frac{1}{4 \, M \, S_{\st}^i} \int \frac{d \xi^-}{2 \pi} \,
 e^{i k \cdot \xi}
 \langle P,S | \; \bar{\psi}(0) \, \gamma^+ i \partial_{\st}^i \gamma_5 \, \psi(\xi) \; 
 | P,S \rangle \,,
\\
\dbar_1(x,x') & = &
 - \frac{g_s \, P^+}{8 \, M \, S_{\st}^i} \int \frac{d \xi^-}{2\pi} \frac{d \eta^-}{2\pi} \,
 e^{i k \cdot \xi - i k' \cdot \eta}
 \langle P,S | \; \bar{\psi}(\eta) \, \gamma^+ \atslash(0)
 \gamma^i \gamma_5 \, \psi(\xi) \; | P,S \rangle \,,
\\
\dbar_2(x',x) & = &
 - \frac{g_s \, P^+}{8 \, M \, S_{\st}^i} \int \frac{d \xi^-}{2\pi} \frac{d \eta^-}{2\pi} \,
 e^{i k' \cdot \eta - i k \cdot \xi}
 \langle P,S | \; \bar{\psi}(\xi) \, \gamma^+ \gamma^i \atslash(0)
 \gamma_5 \, \psi(\eta) \; | P,S \rangle \,,
\end{eqnarray}
while the ones in Eq.~(\ref{rel_2b}) are
\begin{eqnarray}
\widetilde{K}(x) & = & - \frac{1}{4 \, M \, \lambda} \int \frac{d \xi^-}{2 \pi} \,
 e^{i k \cdot \xi}
 \langle P,S | \; \bar{\psi}(0) \, \gamma^+ i \deltslash_{\st} \gamma_5 \,
 \psi(\xi) \; | P,S \rangle \,,
\\
\widetilde{D}_1(x,x') & = &
 - \frac{g_s \, P^+}{8 \, M \, \lambda} \int \frac{d \xi^-}{2\pi} \frac{d \eta^-}{2\pi} \,
 e^{i k \cdot \xi - i k' \cdot \eta}
 \langle P,S | \; \bar{\psi}(\eta) \, \gamma^+ \atslash(0)
 \gamma_5 \, \psi(\xi) \; | P,S \rangle \,,
\\
\widetilde{D}_2(x',x) & = &
 - \frac{g_s \, P^+}{8 \, M \, \lambda} \int \frac{d \xi^-}{2\pi} \frac{d \eta^-}{2\pi} \,
 e^{i k' \cdot \eta - i k \cdot \xi}
 \langle P,S | \; \bar{\psi}(\xi) \, \gamma^+ \atslash(0)
 \gamma_5 \, \psi(\eta) \; | P,S \rangle \,,
\end{eqnarray}
with $\dbar(x,x') = \frac{1}{2} [ \dbar_1(x,x') + \dbar_2(x',x)]$ and 
$\widetilde{D}(x,x') = \frac{1}{2} [ \widetilde{D}_1(x,x') + \widetilde{D}_2(x',x) ]\,$.
In principle, Eqs.~(\ref{rel_1a},\ref{rel_1b}) and Eqs.~(\ref{rel_2a},\ref{rel_2b}) 
(henceforth, referred to as Set-A and Set-B respectively) should contain the same
information and, in fact, assumed to be identical. 
But there exists hardly any proof of that.

These relations are quite remarkable, in particular, since they involve at the same time 
functions describing longitudinally and transversely polarized targets and therefore 
will provide us with a consistency check while comparing data for the measured structure 
functions from different experiments. 
Moreover, they can be quite useful to predict the evolution of one of the structure 
functions once the evolutions of others are known, as has been done 
in Refs.~\cite{bukhvostov_84,belitsky_97a,henneman_01}. 
Keeping their importance in mind, it is worthwhile to delve more into these relations.
We do this in the next section by checking them through explicit calculations 
for a dressed quark target in the framework of light-front time-ordered pQCD.

%%%%%%%%%%%%%%%%%%%%%%%%%%%%%%%%%%%%%%%%%%%%%%%%%%%%%%%%%%%%%%%%%%%%%%%%%%%
\section{Parton Distributions: Results and Discussion} \label{sec_3}
%%%%%%%%%%%%%%%%%%%%%%%%%%%%%%%%%%%%%%%%%%%%%%%%%%%%%%%%%%%%%%%%%%%%%%%%%%%

Before presenting our results, it is useful to disentangle the twist-3 parton distributions
into simpler structures which manifest the different aspects of the QCD dynamics contained 
in them. 
To achieve this, we re-express the structure functions $g_T$ and $h_L$ in terms of 
dynamical fields, the so-called good fields, like in the twist-2 case which right from the 
beginning contains only the good fields.
That is, we eliminate the constrained field $\psi_-$ via the constraint equation 
(\ref{constr}) in terms of $\psi_+$ and $A_{\st}^i$ which are the only dynamical fields 
in the hamiltonian formulation of light-front QCD. 
Thus, $g_T$ and $h_L$ defined in the previous section become
\begin{eqnarray}
  g_T(x,Q^2) &=& {P^+ \over 4 \, M \, S^i_T } \int
		{d\xi^- \over 2\pi} \, e^{i k \cdot \xi} 
		\langle P,S | \Big(\obar_m + \obar_{k_{\st}} 
		+ \obar_g \Big) |P,S \rangle \,, 
		\nonumber \\
		&=& g_T^m(x,Q^2) + g_T^{k_{\st}}(x,Q^2) + g_T^g
			(x,Q^2) \,,  \label{gto} 
                     \vphantom{\frac{1}{1}} \\
	 h_L(x,Q^2) &=& {P^+ \over 8 \, M \, \lambda } \int
		{d\xi^- \over 2\pi} \, e^{ik\cdot \xi} 
		\langle P,S | \Big(\widetilde{O}_m + \widetilde{O}_{k_{\st}} 
		+ \widetilde{O}_g \Big) |P,S \rangle \,,
                \nonumber \\ 
		&=& h_L^m(x,Q^2) + h_L^{k_{\st}}(x,Q^2) + h_L^g
			(x,Q^2) \,,  \vphantom{\frac{1}{1}} \label{hlo} 
\end{eqnarray}
where we have introduced the operators
\begin{eqnarray}
	&& \obar_m=m_q \, \psi_+^\dagger (0) \, \gamma^i 
		\Bigg({1 \over i \roarrow{\partial}^+} - {1\over 
		i \loarrow{\partial}^+} \Bigg) \gamma_5 \, \psi_+(\xi) \,,
			\nonumber \\
	&& \obar_{k_{\st}}= -\psi_+^\dagger (0) \Bigg(\gamma^i
		{1\over\roarrow{\partial}^+} \roarrow{\deltslash}_{\st} 
		+ \loarrow{\deltslash}_{\st} {1\over \loarrow{
		\partial}^+} \gamma^i \Bigg) \gamma_5 \, \psi_+(\xi) \,, 
                \nonumber \\
	&& \obar_g = g_s \, \psi_+^\dagger(0) \Bigg( \atslash(0)
		{1\over i\loarrow{\partial}^+} \gamma^i 
		- \gamma^i {1\over i\roarrow{\partial}^+} \atslash(\xi) 
                \Bigg) \gamma_5 \, \psi_+(\xi) \,, \label{gtdec}
\end{eqnarray} 
and
\begin{eqnarray}
	&& \widetilde{O}_m =2 \, m_q \, \psi_+^\dagger (0)   
		\Bigg({1 \over i \roarrow{\partial}^+} - {1\over 
		i \loarrow{\partial}^+} \Bigg) \gamma_5 \, \psi_+(\xi) \,,
			\nonumber \\
	&& \widetilde{O}_{k_{\st}}= -2 \, \psi_+^\dagger (0) \Bigg(
		{1\over\roarrow{\partial}^+} \roarrow{\deltslash}_{\st}
		+ \loarrow{\deltslash}_{\st} {1\over \loarrow{
		\partial}^+}\Bigg) \gamma_5 \, \psi_+(\xi) \,, 
                \nonumber \\
	&& \widetilde{O}_g = 2 \, g_s \, \psi_+^\dagger(0) \Bigg( \atslash(0)
                {1\over i\loarrow{\partial}^+} 
		-  {1\over i\roarrow{\partial}^+} \atslash (\xi) 
                \Bigg) \gamma_5 \, \psi_+(\xi) \,. \label{hldec}
\end{eqnarray} 
Here $m_q$ is the quark mass and ${\bm A}_{\st} = \sum_a T^a {\bm A}_{\st}^a$ the 
transverse gauge field, while $1 / \partial^+$ is defined in the sense of the principal 
value prescription as given in Eq.~(\ref{pv}). 
The above light-front expressions make the physical picture of twist-3 structure
functions clear. 
It explicitly shows the contributions associated with the quark mass, quark transverse 
momentum and quark-gluon coupling operators. 
Although one naively expects that the contributions depending explicitly on the quark mass 
are suppressed, it turns out that each of them is equally important to extract the information
contained in twist-3 structure functions. 
Notice that Eqs.~(\ref{gto},\ref{hlo}) correspond to what in the literature is often 
referred to as the relations among various light-front correlators coming from the 
QCD equations of motion (see e.g. Ref.~\cite{mulders_96}).

Having presented the relevant issues as far as the operator structures involved in the parton
distributions are concerned, some comments regarding our calculation and the target state 
are in order.
The calculation is straightforward and we shall avoid giving unnecessary details except 
mentioning the following points. 
(For details we refer the reader to Ref.~\cite{harindranath_98}.) 

Firstly, all the required structure functions are calculated for a dressed quark
target given by the following Fock-space expansion truncated at the lowest non-trivial order,
\begin{eqnarray}
	|k , \lambda \rangle &=& {\cal N}\Bigg\{ b^\dagger_\lambda
		(k)|0\rangle + \sum_{\lambda_1\lambda_2} \int 
                {dk_1^+ d^2{\bm k}_{\st 1} \over \sqrt{2 (2\pi)^3 k_1^+}} 
                {dk_2^+ d^2{\bm k}_{\st 2} \over \sqrt{2 (2\pi)^3 k_2^+}} \, 
                \sqrt{2 (2\pi)^3} \; \delta^3(k-k_1-k_2) \nonumber \\
	&&~~~~~~~~~~~~~~~~~~~~~~~~~~~~~~~\times \Phi^\lambda
		_{\lambda_1\lambda_2}(x,{\bm \kappa}_{\st}) \; 
               b^\dagger_{\lambda_1} (k_1) \, a^\dagger_{\lambda_2} (k_2) | 0 \rangle + 
		\cdots\Bigg\} \,, \label{dsqs}
\end{eqnarray}
where $b_\lambda^\dagger (k)$ and $a_\lambda^\dagger (k)$ are the creation operators 
of quarks and gluons on the light-front which obey the usual commutation relations 
(see Eqs.~(\ref{commb},\ref{comma})). 
The most important ingredient in the above dressed quark state is the two particle 
boost-invariant wave-function which can be calculated using light-front time-ordered pQCD 
and is given by
\begin{equation}
	\Phi^\lambda_{\lambda_1\lambda_2}(x, {\bm\kappa}_{\st}) =
             - {g_s \, T^a \over \sqrt{2 (2\pi)^3}} 
               {x \sqrt{1-x} \over {\bm \kappa}_{\st}^2} \,
		\chi^\dagger_{\lambda_1} \Bigg\{ 2\, { \kappa_{\st}^i \over 1-x} 
	        + {1\over x}\, (\tilde{\bm \sigma}_{\st} \cdot {\bm \kappa}_{\st}) \,
		\tilde{\sigma}^i_{\st} - i \, m_q \, \tilde{\sigma}^i_{\st} \,
                {1-x\over x}\Bigg\} \chi_\lambda^{\hphantom{\dagger}} \, 
                \varepsilon^{i*}_{\st,\lambda_2} \,, \label{ap}
\end{equation}
with $x$ and ${\bm \kappa}_{\st}$ being the relative momenta of the quark. 
Note that the $m_q$-dependence in the above wave function has its origin in the helicity 
flip part of the light-front QCD Hamiltonian. 
This is an essential term in investigating the dynamics of transversely polarized targets and, 
hence, is also very important as far as our calculation is concerned. 
The constant ${\cal N}$ appearing in Eq.~(\ref{dsqs}) is determined by the normalization
condition
\begin{equation}
	\langle k' , \lambda' | k , \lambda \rangle
	= 2 (2\pi)^3 \, k^+ \, \delta(k^+ - k'^+) \,
	\delta^2({\bm k}_{\st} - {\bm k}'_{\st}) \, \delta_{\lambda , \lambda'} \,,
\end{equation}
and to the order $\alpha_s$ given by \cite{harindranath_98}
\begin{equation}
 {\cal N} = 1 - {\alpha_s \over 2\pi} \, C_f \ln{Q^2 \over \mu^2}
    \int_0^1 dx \, {1+x^2 \over 1-x} \,.
\end{equation}
Here a hadronic scale $\mu$ has been introduced such that 
${\bm \kappa}_{\st}^2 >> \mu^2 >> (m_q)^2 $, which can be considered as the 
factorization scale separating the ``hard" and ``soft" dynamics of QCD.
This scale $\mu$ also serves as the lower cutoff of the involved transverse mometum 
integration, whereas $Q^2$ is the upper cutoff. 

Secondly, in our calculation we also need a transversely polarized target, for example,
in the case of $g_T$. 
This is obtained by a superposition of two different helicity states. 
Thus, the one polarized in $x$-direction can be expressed by 
\begin{equation}  \label{thb}
	| k , S^1 = \pm 1 \rangle = {1\over \sqrt{2}} 
        \Big(| k , \uparrow \rangle \pm | k , \downarrow \rangle \Big) \,.
\end{equation}

Lastly, the quark mass renormalization enters in the calculation at $\alpha_s$-order and 
we use the following expression for the renormalized quark mass $m_q^R$ in terms of its 
bare mass $m_q$ \cite{wilson_94},
\begin{equation} \label{renq}
	m_q^R = m_q \Bigg( 1 + {3 \, \alpha_s \over 4\pi} \, C_f \ln
		{Q^2\over \mu^2} \Bigg) \,.
\end{equation}

We now present the results of our calculation, i.e., all the relevant structure functions 
for the dressed quark target in Eq.~(\ref{dsqs}) up to order $\alpha_s$.
We first give the twist-3 structure functions $g_T$ and $h_L$.   
It turns out that all the three terms in Eq.~(\ref{gto}) and Eq.~(\ref{hlo}) have nonzero 
contribution to the corresponding twist-3 structure functions and for
clarity we provide them separately. 
For $g_T(x)$ we obtain
\begin{eqnarray}
	g_T^m(x,Q^2) &=& {m_q\over M}
 		\Bigg\{\delta(1-x) + {\alpha_s \over 2\pi} \,
		C_f \ln{Q^2\over \mu^2} \Bigg[{2 \over 
		1-x}  -~\delta(1-x) \int_0^1 dx' \,
		{1+x'^2\over 1-x'} \Bigg]\Bigg\} \,, \label{gtm}\\
	g_T^{k_{\st}}(x,Q^2) &=&- {m_q\over M} \, 
		{\alpha_s \over 2\pi} \, C_f \ln{Q^2\over \mu^2} \, (1-x) \,, 
		\label{gtk} \\
	g_T^g(x,Q^2) &=&  {m_q\over M} \, {\alpha_s 
		\over 2\pi} \, C_f \ln{Q^2\over \mu^2} \, {\delta(1-x) \over 2} \,. 
		\label{gtg}
\end{eqnarray}
A similar calculation for $h_L$ gives
\begin{eqnarray}
	h_L^m(x,Q^2) &=& {m_q\over M}
 		\Bigg\{\delta(1-x) + {\alpha_s \over 2\pi} \, 
		C_f \ln{Q^2\over \mu^2} \Bigg[\, {1\over x}\Bigg({1+x^2 \over 
		1-x}\Bigg) -~\delta(1-x) \int_0^1 dx' \,
		{1+x'^2\over 1-x'} \Bigg]\Bigg\} \,, \label{hlm}\\
	h_L^{k_{\st}}(x,Q^2) &=&- {m_q\over M} \,
		{\alpha_s \over 2\pi} \, C_f \ln{Q^2\over \mu^2} \, {2(1-x)\over x} \,, 
		\label{hlk} \\
	h_L^g(x,Q^2) &=&  {m_q\over M} \, {\alpha_s 
		\over 2\pi} \, C_f \ln{Q^2\over \mu^2}
		\Bigg[ \, {1-x\over x} + {1 \over 2} \, \delta(1-x)\Bigg] \,. 
		\label{hlg}
\end{eqnarray}
Note that the above results represent purely the pQCD dynamics to the $\alpha_s$-order. 
It should be noted that all the individual contributions in $g_T$ as well as $h_L$ 
in the perturbative calculation are of the same order (namely, proportional to $m_q / M$),
which means that the mass dependent terms $g^m_T$ and $h^m_L$ are not suppressed 
contrary to the common belief. 
As mentioned above, $m_q$ is the bare quark mass and up to order $\alpha_s$ it is given 
by Eq.~(\ref{renq}) in terms of the renormalized quark mass $m_q^R$. 
On the other hand, $M$ is the renormalized target mass and, therefore, in our case
it is identical to $m_q^R$ itself, $M=m_q^R$. 
Taking this into account, we finally get $g_T$ and $h_L$ as follows,
\begin{eqnarray}
	g_T(x,Q^2) &=&  
		\delta(1-x) + {\alpha_s \over 2\pi} \, C_f \ln{Q^2\over 
		\mu^2} \Bigg[ \, {1+2x-x^2 \over (1-x)_+} + {1\over 2} \, \delta(1-x) 
		\Bigg] \,, \label{gtf}\\
	h_L(x,Q^2) &=&   
		\delta(1-x) + {\alpha_s \over 2\pi} \, C_f \ln{Q^2\over 
		\mu^2} \Bigg[ \, {2 \over (1-x)_+} + {1\over 2} \, \delta(1-x) 
		\Bigg] \,, \label{hlf}
\end{eqnarray}
where we have used the well-known 'plus'-prescription. 
Eq.~(\ref{gtf}) reproduces\footnote {Our result differs from that obtained
in Ref.~\cite{harindranath_97} by a factor of ${1\over 2}$ which appears in the
definition of $g_T$ that we are using. This is not relevant for our purpose as 
long as we use one consistent set of definitions for all the parton distributions.}
the result already obtained in Ref.~\cite{harindranath_97} for a dressed quark 
target.
Also a covariant one-loop calculation with a quark target yields exactly the same
expression for $g_T$ \cite{altarelli_79}.
Note that the result for $h_L$, obtained for the first time here, does not
contain any singularity at $x = 0$, even though the individual pieces in 
Eqs.~(\ref{hlm}-\ref{hlg}) do.

To investigate the validity of the relations in Set-A, we need to calculate the 
structure functions on the RHS of them which involves explicitly calculating 
$g_1$, $g_{1T}^{(1)}$, $h_1$ and $h_{1L}^{\perp (1)}$ for the dressed quark 
target. 
Carrying out the evaluation of $g_{1T}^{(1)}$ and $h_{1L}^{\perp (1)}$ as given in 
Eq.~(\ref{moments}), we get to the $\alpha_s$-order
\begin{eqnarray}
	g_{1T}^{(1)}(x,Q^2) &=& - 
		{\alpha_s \over 2\pi} \, C_f \ln{Q^2\over \mu^2} \, x(1-x) \,, 
		\label{g1tk} \\
	h_{1L}^{\perp (1)}(x,Q^2)&=&   
		{\alpha_s \over 2\pi} \, C_f \ln{Q^2\over \mu^2} \, (1-x) \,.  
		\label{h1lk} 
\end{eqnarray}
We point out that in the free theory $(\alpha_s = 0)$ these two $k_{\st}$-dependent
functions vanish since our target carries no net transverse momentum.
The same is true for the two functions given earlier in Eqs.~(\ref{gtk},\ref{hlk}).

The results for $g_1$ and $h_1$ to $\alpha_s$-order already exist in the 
literature~\cite{harindranath_98,mukherjee_01} and are given by
\begin{eqnarray}
	g_1(x,Q^2) &=& 
 		\delta(1-x) + {\alpha_s \over 2\pi} \, 
		C_f \ln{Q^2\over \mu^2} \Bigg[\, {1+x^2 \over 
		(1-x)_+} +~{3\over 2} \, \delta(1-x) 
		 \Bigg] \,, \label{g1}\\
	h_1(x,Q^2)&=& 
 		\delta(1-x) + {\alpha_s \over 2\pi} \, 
		C_f \ln{Q^2\over \mu^2} \Bigg[\, {2x \over 
		(1-x)_+}   +~{3\over 2} \, \delta(1-x) 
		 \Bigg] \,. \label{h1}
\end{eqnarray}
Note that the $\alpha_s$-terms for $g_1$ and $h_1$ contain the evolution kernels of the
corresponding structure functions.
Having the explicit results for all the necessary structure functions appearing in the 
Lorentz invariance relations as given in Set-A, we can now compare the LHS and RHS of 
these relations.
By doing so, one readily finds that the relations in Set-A are not satisfied for a 
dressed quark target. 
Therefore, the natural conclusion is either that Lorentz invariance is violated in 
perturbation theory or that the relations in Set-A do not reflect the complete picture. 

It is easy to see that these relations in Set-A in fact do not reflect the complete 
picture. 
To make it evident, we turn our attention now to the relations in Set-B and first 
calculate $\kbar$ and $\widetilde{K}$ for the dressed quark target to the same 
order. 
It turns out that
\begin{equation}
\kbar(x) = g_{1T}^{(1)}(x)~~~ {\rm and}~~~ 
\widetilde{K}(x) = - 2 \, h_{1L}^{\perp (1)}(x) \,.
\end{equation}
This immediately leads us to the conclusion that the relations presented in Set-A 
are actually different from that in Set-B unless the contributions coming from 
the $\dbar(x,x^\prime)$'s and $\widetilde{D}(x,x^\prime)$'s are identically zero, which 
is unlikely in a general scenario. 
Carrying out the explicit calculation of these terms, we get
\begin{eqnarray} \label{dbar}
	&&\dbar_1(x,x^\prime) = \dbar_2(x^\prime,x) = -{\alpha_s \over 2\pi} \, 
		C_f \ln{Q^2\over \mu^2} \, (x^\prime-x) \, \delta(1-x) \,, 
        \\ \label{dtilde}
	&&\widetilde{D}_1(x,x^\prime) = \widetilde{D}_2(x^\prime,x)=
	- {\alpha_s \over 2\pi} \, C_f \ln{Q^2\over \mu^2} \,
	(x^\prime-x) \, \Big[ \delta(1-x) - \delta (1-x^\prime)\Big] \,.
\end{eqnarray}
We point out that no singularity in $(x' - x)$ shows up in 
Eqs.~(\ref{rel_2a},\ref{rel_2b}).
From the above results one easily observes that the apparent pole there gets canceled.

Putting the results in Eqs.~(\ref{g1}-\ref{dtilde}) back in the RHS of 
Eqs.~(\ref{rel_2a},\ref{rel_2b}), we obtain $g_T$ and $h_L$ as given in 
Eqs.~(\ref{gtf},\ref{hlf}) which verifies the relations in Set-B.
Moreover, we see that the discrepancy we found earlier in Set-A is exactly compensated 
by taking these $\dbar(x,x^\prime)$'s and $\widetilde{D}(x,x^\prime)$'s properly 
into account. 
In other words, from this excercise it turns out that the information contained in 
these quark-gluon-quark correlators is missing in the relations given in Set-A, 
thereby making them incomplete.

Therefore, we finally conclude that the relations in Set-A and those in Set-B are not 
identical -- while the first ones are violated, the second ones are fulfilled for a 
dressed quark target up to order $\alpha_s$.
Barring Eqs.~(\ref{gtm}-\ref{gtg}), (\ref{gtf}) and (\ref{g1},\ref{h1}), all the results 
presented in this section are obtained for the first time here in the context of light-front 
QCD.
It should be noted that in the free theory both sets of relations are satisfied, which is easily 
verified by setting all the terms proportional to $\alpha_s$ in the above expressions 
for the structure functions to zero. 
Since only the quark-gluon-quark correlators seem to be missing in Set-A, we believe that
it is valid and useful in models where no gauge fields are involved. 
For instance, we have checked explicitly taking the results for the parton distributions
as obtained in the spectator model~\cite{jakob_97} that the relations in Set-A can be
verified.
However, in the context of a gauge theory like QCD, one should be careful and always
be using the relations in Set-B.

%%%%%%%%%%%%%%%%%%%%%%%%%%%%%%%%%%%%%%%%%%%%%%%%%%%%%%%%%%%%%%%%%%%%%%%%%%%
\section{Sum Rules}\label{sec_4}
%%%%%%%%%%%%%%%%%%%%%%%%%%%%%%%%%%%%%%%%%%%%%%%%%%%%%%%%%%%%%%%%%%%%%%%%%%%

Our calculation here provides a direct way to investigate the existing sum rules for 
twist-3 parton distributions in the case of a dressed quark target.
Defining the structure functions $g_2 \equiv g_T - g_1$ and $h_2 \equiv 2\,(h_L - h_1)$, 
the following sum rules have been proposed in the 
literature \cite{burkhardt_70,burkardt_92,tangerman_94}:
\begin{equation}
\int_0^1 dx~g_2(x) = 0 \,, \qquad \qquad \int_0^1 dx~h_2(x) = 0 \,.
\end{equation}
From the results presented in the previous section (see Eqs.~(\ref{gtf},\ref{hlf}) 
and (\ref{g1},\ref{h1})) we can immediately write down the expressions for $g_2$ and
$h_2$ to the order $\alpha_s$,  
\begin{eqnarray}
	g_2(x,Q^2) &=& {\alpha_s \over 2\pi} \,
		C_f \ln{Q^2\over \mu^2} \Big[ \, 2x 
		-\delta(1-x) 
		 \Big] \, , \label{g2}\\
	h_2(x,Q^2)&=& {\alpha_s \over 2\pi} \, 
		C_f \ln{Q^2\over \mu^2} \Big[ \, 4 -
		 2 \, \delta(1-x) 
		 \Big] \, . \label{h2}
\end{eqnarray}
The BC sum rule for $g_2$ follows readily from Eq.~(\ref{g2}) as has already been shown 
in Ref.~\cite{harindranath_97} using the same method. 
On the other hand, Eq.~(\ref{h2}) gives 
\begin{equation}
\int_0^1 dx~h_2(x) = {\alpha_s \over \pi} \, C_f \ln{Q^2\over \mu^2} \,,
\end{equation}
which shows that the $h_2$ sum rule is violated in perturbation theory.
Incidentally, the second moment of $h_2$ turns out to be zero,
$\int_0^1 dx~x\,h_2(x) = 0$, although the significance of this result is not clear. 

The violation of the $h_2$ sum rule in the context of perturbation theory is a new 
observation.
Such a result is unexpected, bearing in mind that the $h_2$ sum rule has been derived on the 
same footing of rotational invariance like the BC sum rule \cite{burkardt_95}.
We point out that our observation is different from the findings outlined in
Ref.~\cite{burkardt_95}, where a possible violation of this sum rule for the experimentally 
measured structure function has been discussed.
There, the origin of such a violation was attributed to quark zero modes giving rise to 
a $\delta$-function singularity in the parton distribution at $x = 0$.
Since this kinematical point is usually inaccessible, a significant deviation from the 
sum rule could occur in the experiment.
In contrast, we find the violation already at the level of the parton distribution $h_2$
calculated to the order $\alpha_s$. 
Moreover, in our explicit calculation the final result for $h_2$ is not inflicted by 
quark zero modes.

%%%%%%%%%%%%%%%%%%%%%%%%%%%%%%%%%%%%%%%%%%%%%%%%%%%%%%%%%%%%%%%%%%%%%%%%%%%
\section{Conclusions} \label{sec_5}
%%%%%%%%%%%%%%%%%%%%%%%%%%%%%%%%%%%%%%%%%%%%%%%%%%%%%%%%%%%%%%%%%%%%%%%%%%%

In this work we have calculated higher twist and $k_{\st}$-dependent parton distributions
using the light-front hamiltonian description of composite systems in terms of multi-parton 
wave-functions.
Employing a dressed quark target we have evaluated them to the order $\alpha_s$ in 
light-front time-ordered pQCD.
While we have reproduced the results for $g_1$, $h_1$ and $g_T$, all the other results presented
in Sec.~\ref{sec_3} are new. 

These calculations, in particular, have given us the opportunity to study the so-called
Lorentz invariance relations existing among twist-2, twist-3 and $k_{\st}$-dependent
structure functions.
We show explicitly that two distinct sets of such relations exist
in the literature.
While one set is satisfied (Eqs.~(\ref{rel_2a},\ref{rel_2b})) for the dressed quark target, 
the other one (Eqs.~(\ref{rel_1a},\ref{rel_1b})) is not. 
It turns out that quark-gluon-quark correlators are important for the Lorentz invariance
relations, where these pieces are what exactly is missing in Eqs.~(\ref{rel_1a},\ref{rel_1b}). 
The implication of our findings on the existing literature is yet to be explored.

Moreover, we have studied the sum rules for the structure functions $g_2$ and $h_2$.
The BC sum rule for $g_2$ is fulfilled, whereas the corresponding sum rule for $h_2$ is 
violated at the order $\alpha_s$ in perturbation theory.
Since both sum rules have been derived on the same basis of rotational invariance, the
violation of the $h_2$ sum rule is surprising and requires further investigation.

%%%%%%%%%%%%%%%%%%%%%%%%%%%%%%%%%%%%%%%%%%%%%%%%%%%%%%%%%%%%%%%%%%%%%%%%%%%%%%%%%%%%%%%%%
\acknowledgements

We would like to thank A.~Bacchetta, D.~Boer, A.~Henneman and P.~Mulders for useful 
discussions.
This work is part of the research program of the Foundation for Fundamental Research on 
Matter (FOM) and the Netherlands Organization for Scientific Research (NWO) and is 
partially funded by the European Commission IHP program under contract HPRN-CT-2000-00130.

%%%%%%%%%%%%%%%%%%%%%%%%%%%%%%%%%%%%%%%%%%%%%%%%%%%%%%%%%%%%%%%%%%%%%%%%%%%
\appendix
\section{}
In this Appendix we summarize our conventions.
First, we specify the plus and minus lightcone components of a generic 4-vector
$a^{\mu}$ according to $a^{\pm} \equiv a^0 \pm a^3$, and the inner product 
of two 4-vectors is given by 
$a \cdot b = \frac{1}{2} a^+ b^- + \frac{1}{2} a^- b^+ 
 - {\bm a}_{\st} \cdot {\bm b}_{\st} \,.$
For the $\gamma$ matrices we use the light-front 
representation~\cite{harindranath_97}
\begin{equation}  \label{eq_gamma}
\gamma^0 = \left( \begin{array}{rr} 0 & -i \\ 
                                    i & 0 
                  \end{array} \right) , \quad
\gamma^i = \left( \begin{array}{rr} -i \, \tilde{\sigma}_i & 0 \\ 
                                    0 & i \, \tilde{\sigma}_i 
                  \end{array} \right) , \quad
\gamma^3 = \left( \begin{array}{rr} 0 & i \\ 
                                    i & 0 
                  \end{array} \right) , \quad
\gamma^5 = \left( \begin{array}{rr} \sigma_3 & 0 \\ 
                                    0 & -\sigma_3 
                  \end{array} \right) ,
\end{equation}
where $\tilde{\sigma}_1 = \sigma_2$ and $\tilde{\sigma}_2 = - \sigma_1$.
In the usual way we define the dynamical field $\psi_{+} = \Lambda^{+} \psi$ and 
the constrained field $\psi_{-} = \Lambda^{-} \psi$, which follows the constraint equation
\begin{equation} \label{constr}
\psi_{-} = \frac{\gamma^0}{i \partial^{+}} (i \dtslash + m_q) \psi_{+} \,,
\end{equation}
where $D^{\mu} = \partial^{\mu} - i g_s A^{\mu}$ is the covariant derivative.
The operator $1 / \partial^+$ is defined as
\begin{equation} \label{pv}
\frac{1}{\partial^{+}} \, f(x) = \frac{1}{4} \int_{-\infty}^{\infty} dy \;
\epsilon(x - y) \, f(y) \,,
\end{equation}
with $\epsilon(x)$ being the sign-function. 
In the representation (\ref{eq_gamma}), the projection operators
$\Lambda^{\pm} \equiv \gamma^{\mp} \gamma^{\pm} / 4$ take the simple form
\begin{equation}
\Lambda^{+} = \left( \begin{array}{rr} 1 & 0 \\ 
                                    0 & 0 
                  \end{array} \right) , \quad
\Lambda^{-} = \left( \begin{array}{rr} 0 & 0 \\ 
                                    0 & 1  
                  \end{array} \right) .
\end{equation}
For the fermion fields we use the two-component notation~\cite{zhang_93}
\begin{equation}
\psi_{+} = \left( \begin{array}{c} \eta \\ 0 \end{array} \right) , \quad
\psi_{-} = \left( \begin{array}{c} 0 \\
           \frac{1}{i\partial^{+}} [\tilde{\bm \sigma}_{\st} \cdot 
            (i {\bm \partial}_{\st} + g_s {\bm A}_{\st}) + i m_q ] \eta \end{array} \right) \,,
\end{equation}
where the Fourier expansions of the dynamical fields $\eta(x)$ and ${\bm A}_{\st}(x)$ are 
given by
\begin{eqnarray}
\eta(x) & = & \sum_{\lambda} \chi_{\lambda}^{\hphantom{\dagger}} \int 
 \frac{dk^+ \, d^2{\bm k}_{\st}}{2 (2\pi)^3 \sqrt{k^+}}
 \Big( b_{\lambda}(k) \, e^{-i k \cdot x} + d_{-\lambda}^{\dagger}(k) 
 \, e^{i k \cdot x} \Big) \,,
\\
{\bm A}_{\st}(x) & = & \sum_{\lambda} \int \frac{dk^+ \, d^2{\bm k}_{\st}}{2 (2\pi)^3 k^+}
 \Big( a_{\lambda}(k) \, {\bm \varepsilon}_{\st,\lambda} \, e^{-i k \cdot x} 
  + a_{\lambda}^{\dagger}(k) \, {\bm \varepsilon}_{\st,\lambda}^{\ast} 
 \, e^{i k \cdot x} \Big) \,.
\end{eqnarray}
Here the creation and annihilation operators for quarks (gluons) obey the
anticommutation (commutation) relations
\begin{eqnarray} \label{commb}
\Big\{ b_{\lambda}(k),b_{\lambda'}^{\dagger}(k') \Big\}
& = & \Big\{ d_{\lambda}(k),d_{\lambda'}^{\dagger}(k') \Big\}
= 2 (2\pi)^3 \, k^{+} \, \delta(k^{+} - k'^{+}) \, \delta({\bm k}_{\st} - {\bm k'}_{\st}) \, 
 \delta_{\lambda,\lambda'} \,,
\\  \label{comma} 
\Big[ a_{\lambda}(k),a_{\lambda'}^{\dagger}(k') \Big]
& = & 2 (2\pi)^3 \, k^{+} \, \delta(k^{+} - k'^{+}) \, 
 \delta({\bm k}_{\st} - {\bm k}'_{\st}) \, 
 \delta_{\lambda,\lambda'} \,.
\end{eqnarray}

%%%%%%%%%%%%%%%%%%%%%%%%%%%%%%%%%%%%%%%%%%%%%%%%%%%%%
%%%%%%%%%%%%%%%%%%%%%%%%%%%%%%%%%%%%%%%%%%%%%%%%%%%%%

%%%%%%%%%%%%%%%%%%%%%%%%%%%%%%%%%%%%%%%%%%%%%%%%%%%%%%%%%%%%%%%%%%%%%%%%%%%%%
\end{document}